\documentclass[aps,prb,twocolumn,a4paper,floatfix,showpacs,superscriptaddress]{revtex4}
\usepackage{graphicx}
\usepackage{hyperref}
\usepackage{amsmath}
\usepackage{lineno}
\usepackage{verbatim}
\usepackage{amssymb}

\newcommand\rr{\mathbf{r}}
\newcommand\qq{\mathbf{q}}
\newcommand\UU{U_{m_{1} m_{2} m_{3} m_{4}}}
\newcommand\dd{\rm{d}}
\usepackage{color}
\usepackage[normalem]{ulem}

\begin{document}

\title{Doubly screened Coulomb correction approach for strongly correlated systems}

\author{Bei-Lei Liu}
\affiliation{Laboratory of Computational Physics, Institute of Applied Physics and Computational Mathematics, Beijing 100088, China}
\affiliation{School of Mathematical Sciences, Beijing Normal University, Beijing 100875, China}

\author{Yue-Chao Wang\footnote{Corresponding authors: yuechao\_wang@126.com}}
\affiliation{Laboratory of Computational Physics, Institute of Applied Physics and Computational Mathematics, Beijing 100088, China}

\author{Yu Liu\footnote{Corresponding authors: liu\_yu@iapcm.ac.cn}}
\affiliation{Laboratory of Computational Physics, Institute of Applied Physics and Computational Mathematics, Beijing 100088, China}

\author{Hai-Feng Liu}
\affiliation{Laboratory of Computational Physics, Institute of Applied Physics and Computational Mathematics, Beijing 100088, China}

\author{Hai-Feng Song\footnote{Corresponding authors: song\_haifeng@iapcm.ac.cn}}
\affiliation{Laboratory of Computational Physics, Institute of Applied Physics and Computational Mathematics, Beijing 100088, China}

\begin{abstract}
Strongly correlated systems containing $d/f$-electrons present a challenge to conventional density functional theory (DFT), such as the widely used local density approximation (LDA) or generalized gradient approximation (GGA).
In this work, we developed a doubly screened Coulomb correction (DSCC) approach to perform on-site Coulomb interaction correction for strongly correlated materials.
The on-site Coulomb interaction between localized $d/f$-electrons is determined from a model dielectric function that includes both the static dielectric and the Thomas-Fermi screening.
All parameters of the dielectric model are efficiently obtained from self-consistent calculations. We applied DSCC to simulate the electronic and magnetic properties of typical $3d$, $4f$ and $5f$ strongly correlated systems. The results show that the accuracy of DSCC is comparable to hybrid functionals, but an order of magnitude faster. In addition, DSCC can reflect the difference in the Coulomb interaction of the same element between metallic and insulating situations, similar to the popular but computationally expensive constrained random phase approximation (cRPA) approach. This feature suggests that DSCC is also a promising method for simulating Coulomb interaction parameters.
\end{abstract}
\pacs{ }
\maketitle


\section{Introduction}

Materials containing localized $d/f$-electrons exhibit many unique electronic and magnetic properties, and have been used exclusively in a wide variety of applications\cite{Dagotto2005,Gebhard1997,Stewart1984,Moore2009}. Due to the strong electronic correlations involved, the accurate description of these systems is currently considered one of the greatest challenges in $ab ~ initio$ research\cite{Savrasov2001,Carter2008,Jiang2015}. Density functional theory (DFT)\cite{Kohn1965,Hohenberg1964,Martin2004} provides a powerful method for a wide class of real materials, but conventional DFT based on popular local density approximation (LDA)\cite{Kohn1965,Ceperley1980} or generalized gradient approximation (GGA)\cite{Perdew1996-1} fails to qualitatively describe the electronic-structure properties of these strongly correlated systems. For instance, the band gaps of these materials are severely underestimated and ground magnetic states wrongly predicted by LDA/GGA\cite{Cramer2009,Wen2012}. The main deficiency of these local and semilocal functionals is the presence of self-interaction error (SIE), which is especially significant for the highly localized $d$/$f$-electrons. In order to effectively handle strongly correlated systems, amount of effort has been devoted to the development of first-principles approaches.

One class of methods applies an unbiased correction to all valence electrons, such as the GW method \cite{Aryasetiawan1995,Massidda1995,Jiang2009,Jiang2012}, the hybrid functional \cite{Kudin2002, Prodan2005, Silva2007, Chevrier2010, Iori2012}. By partially mixing the Hartree-Fock exact exchange to reduce the SIE, the hybrid functional such as PBE0\cite{Perdew1996-2,Burke1997} and HSE\cite{Heyd2003,Heyd2006} substantially improve the description of properties for many strongly correlated systems\cite{Kudin2002, Prodan2005, Chevrier2010, Iori2012}. The GW method has a more accurate performance for many $3d$ and $4f$ metal oxides\cite{Aryasetiawan1995,Massidda1995,Jiang2009,Jiang2012}. However, it requires much more computational resources than hybrid functionals.
In conventional hybrid functionals, a crucial factor affecting the overall performance of the hybrid functional, the mixing parameter of the exact exchange $\alpha$, is generally assigned a fixed value independent of the material under study. Recently, a class of system-specific mixing parameter hybrid functionals have been developed\cite{Marques2011, Skone2014, Skone2016, Chen2018, Cui2018}. Based on the relationship between hybrid functionals and the static Coulomb hole plus screened exchange (COHSEX) approximation\cite{Hedin1965}, the mixing parameter $\alpha$ is determined by the inverse of the static dielectric constant $\epsilon^{-1}_{\infty}$. For example, Marques et al.\cite{Marques2011} suggested to determine $\epsilon_{\infty}$ at the Perdew–Burke–Ernzerhof (PBE)\cite{Perdew1996-1} level, or from an empirical relation with the estimator based on density gradient. Skone et al.\cite{Skone2014} proposed the dielectric dependent hybrid (DDH) calculated the static dielectric constant $\epsilon_{\infty}$ in a self-consistent way. The range-separated DDH (RS-DDH) scheme\cite{Skone2016}, dielectric-dependent range-separated hybrid functional based on the Coulomb-attenuating method (DD-RSH-CAM)\cite{Chen2018} and doubly screened hybrid functional (DSH)\cite{Cui2018} have also been proposed to improve the performance for narrow-gap systems. It should also be mentioned that DSH highlighted the importance of dielectric screening in the calculation of the mixing parameter of the hybrid functional.

All these newly hybrid functionals generally outperform conventional hybrid functionals in electronic structure calculations for a large class of $sp$ and narrow-gap semiconductors. Nevertheless, Liu et al.\cite{Liu2020} recently found that two advanced non-empirical hybrid functionals, DD-RSH-CAM and DSH, severely overestimate the band gaps of transition-metal monoxides, this suggests that such hybrid functionals with self-consistent mixing parameter may be theoretically unsatisfactory in strongly correlated systems. Moreover, the calculation of the exact exchange energy in the hybrid functional is demanding for the extended system, even for short-range hybrid functionals such as HSE.


Another class of methods imposes a local correction term only on the correlated electrons (typically, e.g., $d/f$ electrons). One of the most popular implementations introduces on-site Coulomb and exchange interactions explicitly into Hamiltonian through simplified Hubbard model, such as DFT+X [X = Hubbard $U$ correction\cite{Anisimov1991,Czyzyk1994,Dudarev1998,Himmetoglu2014} ($U$), dynamical mean field theory\cite{Lichtenstein1998,Kotliar2006,Anisimov2010} (DMFT), or Gutzwiller approximation\cite{Deng2008,Lanata2015}]. DFT+$U$ is currently the most widely used method in DFT+X, since its efficiency. However, the performance of these methods highly relies on the values of the on-site interaction parameters (Hubbard $U$ and Hund $J$)\cite{Anisimov2010, Himmetoglu2014}.
We also note that the method “exact exchange for correlated electrons” (EECE) proposed by Novak et al.\cite{Novak2006}, employs a Hartree-Fock exchange term similar to that in the hybrid functionals, but only for the correlated electron subspace.

In practice, the $U/J$ value is usually considered as an empirical parameter, or derived from the semi-empirical tuning to experimental data\cite{Bengone2000,Castleton2007,Pegg2017}, which reduces the predictive power of the DFT+X. Several first-principles approaches to calculating $U/J$ do existed. The constrained DFT (cDFT) method\cite{Dederichs1984} calculate $U$ by the energy difference of different electronic configurations in the full-potential framework. The linear-response (LR) formalism\cite{Cococcioni2005} extends cDFT to a more efficient pseudopotential framework, calculating the $U$-value by perturbation technique. The constrained random phase approximation (cRPA)\cite{Aryasetiawan2004} based on many-body perturbation theory, which is capable of providing frequency-dependent screened Coulomb interactions, is also a promising method. Nevertheless, these approaches can be computationally demanding, especially for complex materials or in large cells. By evaluating the screening parameter, the local screened Coulomb correction (LSCC) method\cite{Wang2019} has been proved efficient in many strongly correlated systems. However, LSCC only considers the Thomas-Fermi screening in the metallic system. It does not take into account the dielectric screening that plays a dominant role in gapped systems.

In this work, we use a model dielectric function to determine the on-site interactions between correlated electrons, proposed the doubly screened Coulomb correction (DSCC) approach. The dielectric model considers the influence of both dielectric and metallic screenings, which are dominant in the gapped systems and metals, respectively. We assess the performance of DSCC in the simulation of electronic-structure properties of a range of strongly correlated systems (MnO, FeO, CoO, NiO, CeO$_2$, Ce$_2$O$_3$, EuO, EuS, UO$_2$, NpO$_2$, PuO$_2$, and AmO$_2$), including band gaps and relative magnetic stability, and compared with hybrid functionals, DFT+$U$ and LSCC. The performance of DSCC in the calculation of on-site correlation strength is also investigated.

The paper is organized as follows. In Section \ref{sec:Meth}, we introduce the theory of the DSCC approach, describe the implementation and present computational details. In Section \ref{sec:Res}, we show the performance of DSCC in typical strongly correlated systems. Finally, we summarizes the work of this paper in Section \ref{sec:Con}.


\section{Methodology}
\label{sec:Meth}

\subsection{Doubly Screened Coulomb Correction approach}

In the DSCC approach, an on-site interaction model is used to correct correlated $d/f$-electrons on the basis of the DFT.
Given screened Coulomb potential $v_{\rm sc}$ and local orbitals $\left\{\varphi_m\right\}$, the interaction energy of these correlated electrons $E_{\rm on-site}$ takes the form\cite{Anisimov2010}
\begin{equation}
\begin{split}
\label{eq:on-site}
&E_{\rm on-site} =\frac{1}{2}\sum_{{m},\sigma}\big[\UU n_{m_1,m_2}^\sigma n_{m_3,m_4}^{\bar{\sigma}}  \\
& +(\UU-U_{m_{1} m_{3}m_{4}m_{2}})n_{m_1,m_2}^\sigma n_{m_3,m_4}^\sigma\big],
\end{split}
\end{equation}
where $\left\{m\right\}$ is the index set of local orbitals, $m_1,m_2,m_3,m_4\in\{m\}$, $\sigma$ stands for the spin,
\begin{equation}
\label{eq:int}
U_{ijkl}=\int\int{\varphi_i^\ast(\rr)\varphi_j^\ast(\rr')v_{\rm sc}(\rr,\rr')\varphi_k{(\rr)\varphi}_l(\rr')\dd\rr \dd\rr'}
\end{equation}
is the elements of the interaction matrix, and $n_{i,j}^\sigma$ is the elements of the occupation matrix, which is determined by the projection of Kohn-Sham orbitals onto the subspace spanned by local orbitals\cite{Shick1999,Bengone2000,Amadon2008}. The screened Coulomb potential $v_{\rm sc}$ could be calculated by the dielectric function $\epsilon$, as
\begin{equation*}
v_{\rm sc}(\rr,\rr')=\int_{\mathbb{R}^3}{\epsilon^{-1}(\rr,\rr'')}v(\rr'',\rr')\dd\rr'',
\end{equation*}
where $v$ is the Coulomb potential. In this work, we take the following model dielectric function to construct the screened Coulomb potential\cite{Cappellini1993},
\begin{equation}
\label{eq:eps}
\epsilon(\qq) = 1+ \left\{(\epsilon_\infty - 1)^{-1}+\left(\frac{q}{\lambda_{\rm TF}}\right)^2\right\}^{-1},
\end{equation}
where $\qq$ being wave vector in the cell of reciprocal space, $q = |\qq|$, $\epsilon_\infty$ is the static
dielectric constant, and $\lambda_{\rm TF}$ is the screening parameter in the conventional Thomas-Fermi screening model\cite{Wang2019}. Using the transformation from reciprocal to real space of $\epsilon(\qq)$, the corresponding screened Coulomb potential is
\begin{equation}
\label{eq:vsc}
v_{\rm sc}(\rr,\rr')=\frac{1}{\epsilon_\infty}\frac{1}{|\rr-\rr'|}+\left(1-\frac{1}{\epsilon_\infty}\right)\frac{e^{-{\bar{\lambda}}_{\rm TF}|\rr-\rr'|}}{|\rr-\rr'|},
\end{equation}
where ${\bar{\lambda}}_{\rm TF}=\epsilon_\infty (\epsilon_\infty-1)^{-1} \lambda_{\rm TF}$.
Physically, the dielectric function Eq.\ref{eq:eps} captures both the static dielectric screening, which is important in the gapped system, and the Thomas-Fermi screening in the metallic systems. In the limit of the ideal metal, $\epsilon_\infty\rightarrow\infty$, it takes the form of the dielectric function in the Thomas-Fermi screening model
\begin{equation*}
\epsilon(\qq) = 1+{(\lambda_{\rm TF}/q)}^2,
\end{equation*}
which performed the same as LSCC.

The total energy of DSCC can be written as follow:
\begin{equation*}
E_{\rm tot}=E_{\rm DFT}[\rho]+E_{\rm on-site}[\{\varphi_m\},v_{\rm sc}]-E_{\rm dc},
\end{equation*}
where $E_{\rm DFT}$ refers to the energy of the whole system described by DFT. The double counting term $E_{\rm dc}$ is  used to cancel the correlated electron-electron interaction that have already been partially included in DFT. The elements of the interaction matrix Eq.\ref{eq:int} can be obtained by screened Slater integrals $F_n$ and $a_n(i,j,k,l)$, which can be calculated by Gaunt coefficients\cite{Amadon2008,Wang2019},
\begin{equation}
U_{ijkl}=\sum_{n=0}^{2l} a_n(i,j,k,l) F_n.
\end{equation}
We explicitly calculate the screened Slater integrals $F_n$ by
\begin{equation*}
\begin{split}
F_{n}&=-(\frac{\epsilon_\infty-1}{\epsilon_\infty})(2n+1){\bar{\lambda}}_{\rm TF} \\
&\times\iint R_l^2(r)j_n(i{\bar{\lambda}}_{\rm TF}r_<)h_n(i{\bar{\lambda}}_{\rm TF}r_>)R_l^2(r')r^2{r'}^2\dd r \dd r' \\
&+ \frac{1}{\epsilon_\infty}\iint R_l^2(r)R_l^2(r')r^2{r'}^2\frac{r_<^n}{r_>^{n+1}}\dd r\dd r'.
\end{split}
\end{equation*}
using Eq.\ref{eq:vsc}, where $j_n$ and $h_n$ are the spherical Bessel function and the spherical Hankel function of the first kind, respectively. $r_<=\min\{r,r'\}$, and $ r_>=\max\{r,r'\}$. $R_l(r)$ is the radial part of the atomic-type local orbitals $\varphi_m$, $\varphi_m(r) = R_l(r)Y_{lm}(\hat{r})$, where $l$ is the orbital angular momentum quantum number. In particular, the Hubbard $U$ and Hund $J$ is related to the screened Slater integrals $F_n$ as follows\cite{Anisimov2010,Norman1995}:
\begin{equation}
\begin{split}
\label{eq:uj}
U &= F_{0}, \\
J &= \frac{F_{2}+F_{4}}{14}\ (l=2), \\
&= \frac{286F_{2}+195F_{4}+250F_{6}}{6435}\ (l=3).
\end{split}
\end{equation}
Note that in DFT+$U$, the elements of the interaction matrix are not calculated explicitly, but by fixing the ratios between $F_n$s and then a linear combination of the input parameters $U$ and $J$\cite{Liechtenstein1995}.

\subsection{Calculation of parameters and implementation}

A key issue in using the model dielectric function is how to determine the parameters $\lambda_{\rm TF}$ and $\epsilon_\infty$. The screening parameter $\lambda_{\rm TF}$ could be obtained by the electron density of system. We calculated $\lambda_{\rm TF}$ in a weighted averaging way following the work Wang and Jiang (see S-III type)\cite{Wang2019}.  The static dielectric constant $\epsilon_\infty$ is calculated using the density functional perturbation theory (DFPT)\cite{Gajdos2006} (i.e., coupled-perturbed Kohn-Sham, CPKS method\cite{Ferrero2008}).

\begin{figure}[htbp]
\centering
\includegraphics[width=0.44\textwidth]{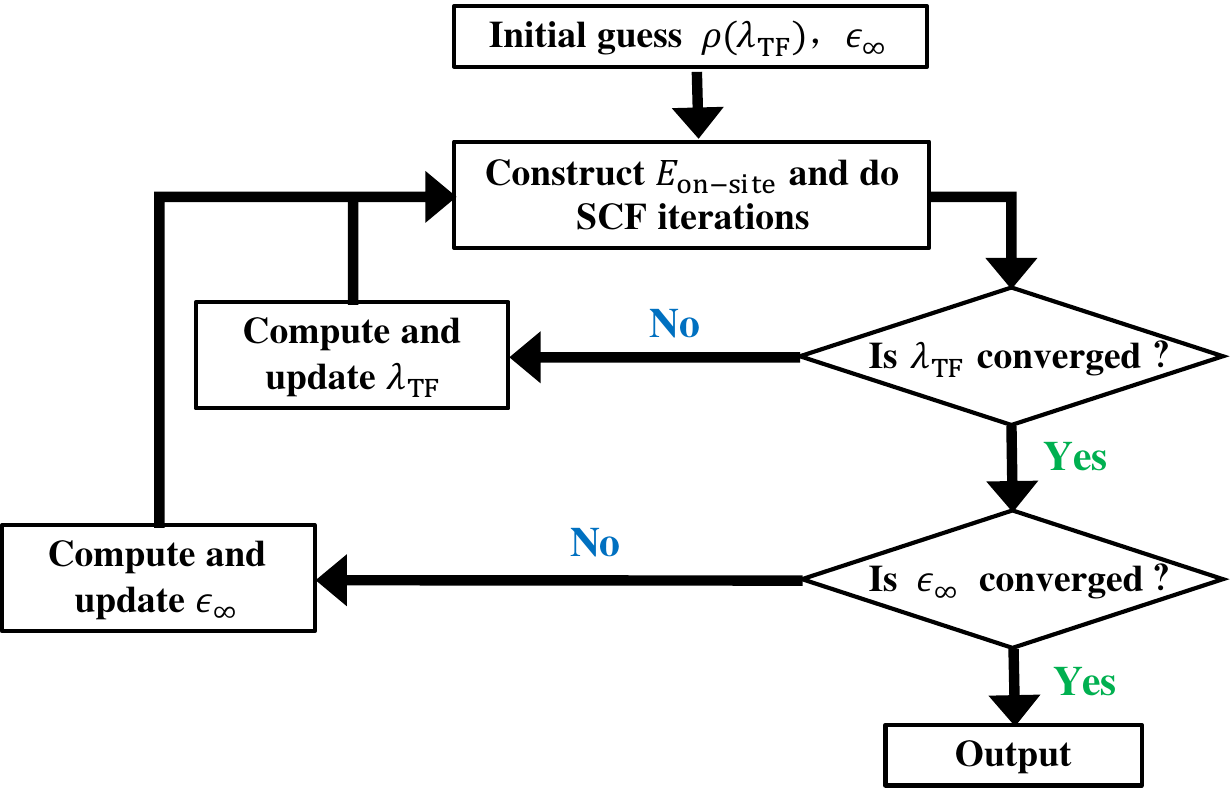}
\caption{Workflow of self-consistent DSCC approach. A one-shot calculation from converged $\lambda_{\rm TF}$ also be applied in this work, which is denoted as DSCC$_0$.}\label{fig:self}
\end{figure}

The self-cosistent cycle of DSCC is illustreated in Fig.\ref{fig:self}. We used an initial guess of $\lambda_{\rm TF}$ according to its relationship with electron density $\rho$, and used an initial $\epsilon_\infty$ from the LSCC calculation. The calculation of $\lambda_{\rm TF}$ is rather simple than $\epsilon_\infty$. Thus, we self-consistently update $\lambda_{\rm TF}$ and $\epsilon_\infty$ separately. Besides the self-consistent DSCC, we also consider the scheme that uses self-consistent $\lambda_{\rm TF}$ , but $\epsilon_\infty$ without further updating. We denote it as DSCC$_0$ hereafter.

In this work, the radial part of local orbital is constructed as\cite{Novak2006}
\begin{equation}
\varphi_{l}(r) = \sqrt{\frac{\rho_{l}(r)}{N_l}},
\end{equation}
where $\rho_{l}(r)$ is the spherically symmetric part of $l$-electron density within the projector augmented wave (PAW)\cite{Blochl1994,Kresse1999} augmentation region, and $N_l$ is the occupation number of the corresponding orbitals. To avoid empirical tuning and to achieve good performance at low computational cost by considering different screening effects, we use a model dielectric function with a simple form and few parameters. It should be noted that there are also more complex models of the dielectric function, like the form\cite{Cappellini1993,Palummo1995}:
\begin{equation*}
\epsilon(\qq)=1+\left[ {(\epsilon_\infty - 1)}^{-1} + \alpha\left(\frac{q}{\lambda_{\rm TF}}\right)^2+\frac{3q^4}{4 k_{\rm F}^2 \lambda_{\rm TF}^2}\right]^{-1},
\end{equation*}
where $\alpha$ is an empirical parameter, $k_{\rm F}$ is the Fermi wavevector. For $q \rightarrow \infty$, it reproduces the behavior of free-electron gas.

\subsection{Computational details}

We have implemented the DSCC approach in Vienna ab initio simulation package (VASP)\cite{Kresse1993, Kresse1996}.
All the calculations are performed in the local modified version of VASP code. We study the performance of DSCC approach in a range of $d/f$-electron systems, i.e., TMO (TM = Mn, Fe, Co, Ni), CeO$_2$, Ce$_2$O$_3$, EuX (X = O, S), and AnO$_2$ (An = U, Np, Pu, Am). The Mn\_sv, Fe\_sv, Co\_sv, Ni\_
pv and standard Ce, Eu, U, Np, Pu, Am, O, and S PAW\cite{Blochl1994,Kresse1999} potentials are used, and the PBE\cite{Perdew1996-1} exchange-correlation functional is applied for all calculations. We adopted the fully localized limit (FLL) version\cite{Anisimov1993}, one of the most popular schemes to define the double counting term $E_{\rm dc}$. All the plane-wave energy cutoffs are set as 600 eV. A $\Gamma$-centered ${\bf k}$-mesh of $6\times6\times6$ is applied to ensure convergence, and for metals, a finer $15\times15\times15$ $\bf k$-mesh is applied. The energy convergence criterion is chosen to be $10^{-6}$ eV for self-consistent field (SCF) iterations. In order to ensure the convergence of the ground state, an occupation matrix control scheme is used for the $d/f$-orbitals\cite{Allen2014,Chen2022}.
For TMO and EuX, we consider the type-II antiferromagnetic (AFM) ordering along the (111) direction\cite{Shi2008,Terakura1984}. For Ce$_2$O$_3$, we consider the AFM phase that two Ce atoms in the hexagonal unit cell with antiparallel magnetic momenta. For AnO$_2$ we consider the AFM ordering along the (001) direction\cite{Wen2012}. We use experimental crystal structures, taken from Ref. \cite{Tran2006} for TMO, Ref. \cite{Silva2007} for CeO$_2$, and Ce$_2$O$_3$, Ref. \cite{Shi2008} for EuX, and Ref. \cite{Moree2021} for AnO$_2$, in all calculations.


\section{Results and Discussion}
\label{sec:Res}

\subsection{Electronic energy gaps}

\begin{table*}[tbp]
\caption{\label{tab:gap} The band gap (in unit of eV) calculated by DSCC$_0$, sc-DSCC and other approaches are compared with the experimental electronic gaps for typical strongly correlated systems. Except for EuO, EuS, and CeO$_2$, all systems are in the AFM phase. EuO and EuS are in the FM phase and CeO$_2$ is in the nonmagnetic phase. The last three rows show the mean error (ME), the mean absolute error (MAE) and mean absolute relative error (MARE) of the results from different methods with respect to the experimental data.}
\begin{tabular*}{0.95\textwidth}{@{\extracolsep{\fill}} l|cccccccc}
\hline\hline
System     &PBE      &PBE0                    &HSE03                   &PBE+$U$                            &LSCC  &DSCC$_{0}$  &sc-DSCC  &Exp.\\
 \hline
MnO        &0.86     &3.65\cite{Liu2020}      &2.65\cite{Liu2020}      &2.53\cite{Shih2012}                &1.59  &2.11        &2.17    &3.9\cite{vanElp1991-1}\\
FeO        &Metal    &3.02\cite{Liu2020}      &2.11\cite{Liu2020}      &1.88\cite{Shih2012}                &0.59  &2.02        &2.08    &2.4\cite{Parmigiani1999}\\
CoO        &Metal    &4.25\cite{Liu2020}      &3.21\cite{Liu2020}      &2.99\cite{Shih2012}                &2.34  &2.83        &2.89    &2.5\cite{vanElp1991-2}\\
NiO        &0.95     &5.29\cite{Liu2020}      &4.28\cite{Liu2020}      &3.02\cite{Shih2012}                &3.19  &3.41        &3.47    &4.0\cite{Hufner1984}\\
CeO$_2$    &1.95     &4.50\cite{Silva2007}    &3.50\cite{Silva2007}    &2.50\cite{Silva2007}               &2.18  &2.33        &2.34    &$\sim$3\cite{Wuilloud1984}\\
Ce$_2$O$_3$ &0.32    &3.50\cite{Silva2007}    &2.50\cite{Silva2007}    &2.60\cite{Silva2007}               &2.03  &2.70        &2.73    &2.4\cite{Prokoviev1996}\\
EuO        &Metal    &1.02\cite{Schlipf2013}  &0.23\cite{Schlipf2013}  &0.77\cite{Shi2008}                 &0.53  &1.14        &1.26    &1.12\cite{Mauger1986}\\
EuS        &Metal    &1.43\cite{Schlipf2013}  &0.74\cite{Schlipf2013}  &1.33\cite{Shi2008}                 &0.93  &1.40        &1.47    &1.65\cite{Mauger1986}\\
UO$_2$     &Metal    &3.34                    &2.40                    &2.6\cite{Moree2021}                &1.22  &2.36        &2.40    &2.1\cite{Kern1999}\\
NpO$_2$    &Metal    &3.53                    &2.39                    &3.1\cite{Moree2021}                &2.04  &2.77        &2.83    &2.85\cite{McCleskey2013}\\
PuO$_2$    &Metal    &3.24                    &2.51                    &2.2\cite{Moree2021}                &1.67  &2.49        &1.54    &2.80\cite{McCleskey2013}\\
AmO$_2$    &Metal    &2.93                    &1.83                    &0.8\cite{Moree2021}                &0.20  &1.40        &1.44    &1.3\cite{Suzuki2012}\\
ME         &$-$2.16  &0.81                    &$-$0.14                 &$-$0.31                            &$-$0.96 &$-$0.26   &$-$0.20 &---\\
MAE        &2.16     &0.90                    &0.54                    &0.55                               &0.96  &0.42        &0.42    &---\\
MARE [\%] &90        &40                      &26                      &22                                 &41    &15          &15      &---\\
\hline\hline
\end{tabular*}
\end{table*}

\begin{figure}[htbp]
\centering
\includegraphics[width=0.44\textwidth]{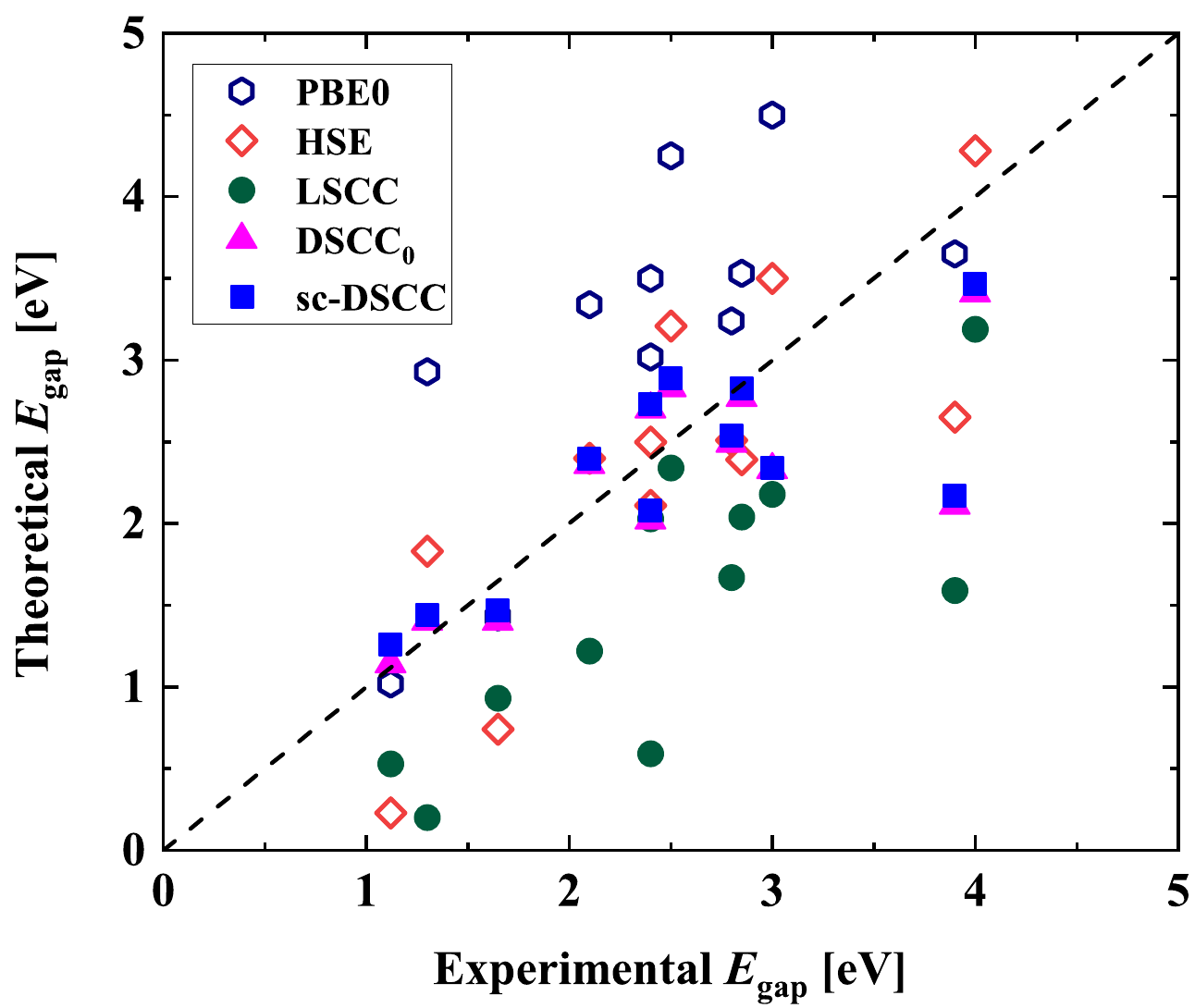}
\caption{Comparison between calculated and experimental band gap ($E_{\rm gap}$) for the set of 12 typical strongly correlated systems listed in Table \ref{tab:gap}.}\label{fig:gap}
\end{figure}

Table \ref{tab:gap} shows the band gaps ($E_{\rm gap}$) derived by the DSCC approach together with the theoretical and experimental data. As reported in previous research, PBE significantly underestimates the band gaps and even yields metallic behavior for many insulators. PBE0 substantially improves the description of band gaps by mixing a fraction of the Hartree-Fock exact exchange. However, it overestimates the values for most systems. The overestimation of PBE0 can be alleviate to some extent by the HSE approach, which provides quite accurate results. PBE+$U$ can also improve the PBE description for strongly correlated systems, but the performance is heavily dependent on the parameter $U$. Using the $U$ values obtained by the first-principles cRPA or LR method, the mean absolute error (MAE) is now reduced to 0.54 eV. The LSCC approach is able to correctly reproduce the insulator behavior of all systems at a computational cost comparable to PBE. Nevertheless, due to the neglect of the important static dielectric screening effect, the band gap from LSCC is underestimated by 0.96 eV in terms of the MAE. Both the one-shot (DSCC$_0$) and self-consistent (sc-DSCC) schemes yield results in good agreement with experiment, archiving a remarkable improvement over LSCC. Moreover, the difference between the one-shot and self-consistent results is less than 0.12 eV. This indicates that DSCC$_0$ can be used as a good correction of LSCC, which is very promising for high-throughput applications. We also illustrate the accuracy of LSCC and DSCC compared to experiment in Fig. \ref{fig:gap} more clearly.

\begin{figure*}[htbp]
\centering
\includegraphics[width=0.95\textwidth]{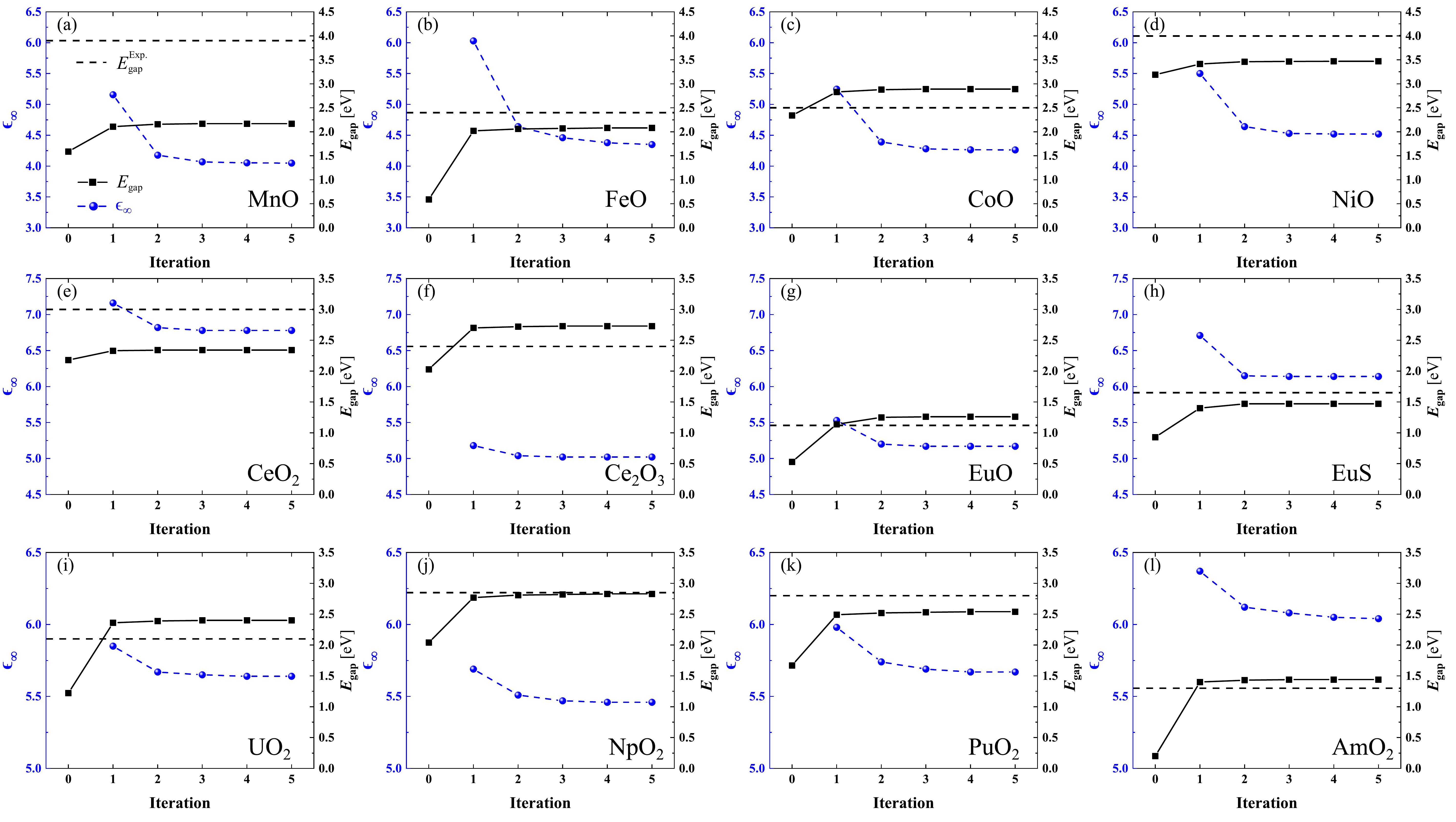}
\caption{Convergence of the value of the static $\epsilon_\infty$ and the band gap $E_{\rm gap}$ in the self-consistent DSCC scheme. The iterative procedure starts from LSCC.}\label{fig:iter}
\end{figure*}

Fig. \ref{fig:iter} illustrate the iterative procedure of calculated $\epsilon_\infty$ and band gap of the DSCC. As shown in Fig. 3, the calculated $\epsilon_\infty$ start from converged $\lambda_{\rm TF}$   generally reach convergence (0.01) in 2-3 steps. Capturing the important static dielectric screening effect in insulators, the band gap can also be converged rapidly. Remarkably, the main computational bottleneck of DSCC lies in the DFPT simulation of the static dielectric constant in each iteration. The rapid convergence enables DSCC to achieve a generally 8-10 times faster speed compared to conventional hybrid functionals.

\subsection{Energies differences of magnetic states}

\begin{table*}[tbp]
\caption{\label{tab:afm-fm}
The energy difference between AFM and FM per formula unit (in meV) calculated by DSCC (self-consistent scheme) and other approaches are compared with the experimental energy difference. The last three rows show the mean error (ME), the mean absolute error (MAE) and mean absolute relative error (MARE) of the results from different methods with respect to the experimental data. Due to the lack of accurate experimental data for Ce$_2$O$_3$ and AnO$_2$ (An = U, Np, Pu, Am), the ME, MAE and MARE derived with respect to the experimental values for else systems. For TMO (TM = Mn, Fe, Co, Ni), PBE+$U$ approach use $U/J$ values calculated by Shih et al. using cRPA \cite{Shih2012}. For CeO$_2$ and Ce$_2$O$_3$, $U/J$ values calculated by Fabris et al. using LR \cite{Fabris2005}. For EuX (X = O and S), $U/J$ values calculated by Shi et al. using LR \cite{Shi2008}. For AnO$_2$, $U/J$ values calculated by Mor\'{e}e, et al. using cRPA \cite{Moree2021}.}
\begin{tabular*}{0.95\textwidth}{@{\extracolsep{\fill}} l|ccccccccc}
\hline\hline
System     &PBE      &PBE0     &HSE03  &PBE+$U$ &LSCC  &DSCC        &Exp.\\
 \hline
MnO        &$-$148.8 &$-$87.1  &$-$95.8  &$-$78.5              &$-$97.8  &$-$49.7        &$-$62\cite{Pepy1974} \\
FeO        &32.0     &$-$29.6  &$-$19.2  &$-$24.4              &$-$30.4  &$-$18.3        &$-$22\cite{Kugel1978}\\
CoO        &$-$141.4 &$-$47.5  &$-$82.6  &$-$58.2              &$-$15.0  &$-$24.5        &$-$34\cite{Tomiyasu2006}\\
NiO        &$-$265.2 &$-$105.4 &$-$130.7 &$-$104.5             &$-$120.3 &$-$90.6        &$-$112\cite{Shanker1973}\\
Ce$_2$O$_3$  &98.8 &$-$4.1 &$-$4.1 &$-$2.0             &$-$3.0 &$-$1.4        &$<$0\cite{Pinto1982}\\
EuO       &49.9 &19.0 &20.5 &13.4             &15.5 &8.2        &8.7\cite{Nielsen1976}\\
EuS       &28.7 &2.3 &3.2 &$-$0.23             &0.65 &$-$2.3        &1.4\cite{Nielsen1976}\\
UO$_2$     &153.8    &82.7     &$-$175.4 &$-$563.8             &$-$204.9 &$-$695.2       &$<$0\cite{Osborn1988}\\
NpO$_2$    &154.6    &7.8      &4.6      &1.9                  &3.1      &1.4            &$<$0\cite{Kern1988}\\
PuO$_2$    &204.5    &22.4     &$-$135.8 &$-$1262.3            &$-$648.0 &$-$1573.6      &$\leq$ 0\cite{Santini1999,Colarieti2002}\\
AmO$_2$    &143.1    &16.6     &23       &$-$13.1              &128.7    &$-$55.8        &$<$0\cite{Tokunaga2010}\\
ME         &$-$37.5  &$-$4.7  &$-$14.1  &$-$5.4              &$-$4.6  &7.1            &---\\
MAE        &78.3     &10.7     &19.6     &9.5                 &13.2     &8.5           &---\\
MARE [\%] &544      &51       &82       &48    &48       &59             &---\\
\hline\hline
\end{tabular*}
\end{table*}

We further investigate the performance of DSCC on the more subtle energy difference between different magnetic orders. Table. \ref {tab:afm-fm} compares the energy differences of AFM phase and ferromagnetic (FM) phase ($E_{\rm AFM-FM}$) calculated with the present DSCC approach and with PBE, PBE0, HSE03, PBE+$U$ ($U$/$J$ values from cRPA) and LSCC. The experimental values are obtained from experimentally determined magnetic coupling constants. TMO, Ce$_2$O$_3$, UO$_2$, NpO$_2$ and AmO$_2$ have AFM ground states\cite{Pepy1974,Kugel1978,Tomiyasu2006,Shanker1973,Pinto1982,Osborn1988,Kern1988,Tokunaga2010}. EuO and EuS have FM ground states\cite{Nielsen1976}. While PuO$_2$ is definitely not FM, its magnetic state is still controversial. It appears to be paramagnetic\cite{Santini1999}, but some authors are of the opinion that it could have AFM exchange\cite{Santini1999,Colarieti2002}. Due to the lack of experimental measurements, accurate energy differences of Ce$_2$O$_3$ and the actinide dioxides have not yet been obtained.

For FeO, Ce$_2$O$_3$ and all actinide dioxides considered, PBE lead to wrong prediction of magnetic state. PBE0 achieved good agreement in considered $3d$- and $4f$- compounds, but yield qualitatively wrong description of actinide dioxides. HSE03 correctly predicts the relative stability of the FM and AFM phases for UO$_2$ and PuO$_2$. The methods based on the correction of localized electrons, such as PBE+$U$, LSCC and DSCC, tend to yield larger energy differences for UO$_2$ and PuO$_2$ than HSE03. LSCC gives a correct description of the relative stability of the magnetic states for all systems except NpO$_2$. Compared with existing experimental data, DSCC could give a relatively quantitative description of considered systems. However, DSCC wrongly predicts an AFM phase for EuS with several meV. This issue also leads to a larger MARE for DSCC, since \% error for this difference is very high.

\subsection{Effective Coulomb interaction strength}

\begin{figure*}[htbp]
\centering
\includegraphics[width=0.9\textwidth]{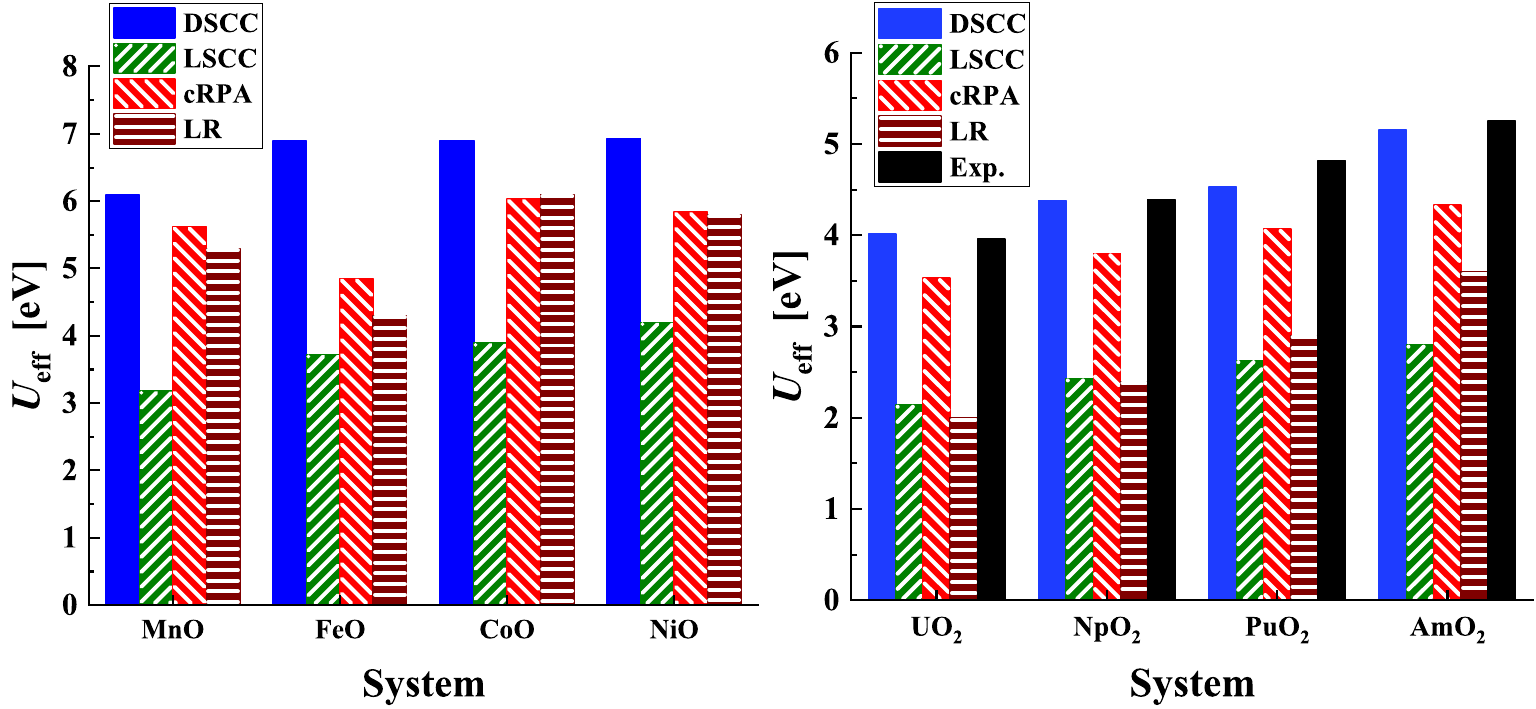}
\caption{The effective Coulomb interactions $U_{\rm eff} \equiv U-J$ obtained from DSCC, LSCC, cRPA, LR and experimental approaches. For TMO (TM = Mn, Fe, Co, Ni), $U^{\rm cRPA}_{\rm eff}$ derived by Shih et al. \cite{Shih2012}, $U^{\rm LR}_{\rm eff}$ derived by Floris et al. \cite{Floris2011}. The experimental data is lacking for TMO. For AnO$_2$ (An = U, Np, Pu, Am), $U^{\rm cRPA}_{\rm eff}$ derived by Mor\'{e}e, et al. \cite{Moree2021}, $U^{\rm LR}_{\rm eff}$ derived by Beridze et al. \cite{Beridze2016}, $U^{\rm Exp.}_{\rm eff}$ derived by Kotani et al.\cite{Kotani1992}.}\label{fig:ueff}
\end{figure*}

Based on the model dielectric function, DSCC can evaluate the on-site electron interaction $U/J$ as Eq.\ref{eq:uj}. Fig. \ref{fig:ueff} shows the effective Coulomb interaction $U_{\rm eff} \equiv U-J$ values obtained by DSCC, LSCC, cRPA, LR and experiment for two series TMO and AnO$_2$. The experimental values are estimated from the experimental photoemission spectra. In the following, we denote the $U_{\rm eff}$ calculated by the DSCC (or LSCC, cRPA, LR, and experiment) approach as $U^{\rm DSCC}_{\rm eff}$ (or $U^{\rm LSCC}_{\rm eff}$, $U^{\rm cRPA}_{\rm eff}$, $U^{\rm LR}_{\rm eff}$, and $U^{\rm Exp. }_{\rm eff}$).

Several features are noteworthy from Fig.\ref{fig:ueff}. For transition-metal monoxides, $U^{\rm DSCC }_{\rm eff}$ and $U^{\rm LSCC }_{\rm eff}$ increase with atomic number, while $U^{\rm cRPA }_{\rm eff}$ and $U^{\rm LR }_{\rm eff}$ have a valley in FeO. For actinide dioxides, all $U_{\rm eff}$ show increasing trends along the $5f$ series. For both series, $U^{\rm DSCC}_{\rm eff}$ are higher than $U_{\rm eff}$ from the other three theoretical approaches. We found that $U^{\rm DSCC}_{\rm eff}$ are closest to the values $U^{\rm Exp.}_{\rm eff}$ estimated from the experimental data (less than 0.3 eV), for actinide dioxides. In addition, as recent studies have shown, $U_{\rm eff} \simeq$ 4.0 eV for UO$_2$ and 5.25 eV for AmO$_2$ provide good simulations of the structural properties, electronic structure properties\cite{Dorado2009, Wang2013, Chen2022, Noutack2019}. While there is a lack of experimental data for TMO, we noted that $U_{\rm eff}$ = 6.0-7.0 eV\cite{Ren2006, Kunes2007, Kunes2008, Korotin2008} used in DFT+DMFT simulations lead to accurate performance in the electronic structure.

\begin{figure*}[htbp]
\centering
\includegraphics[width=0.9\textwidth]{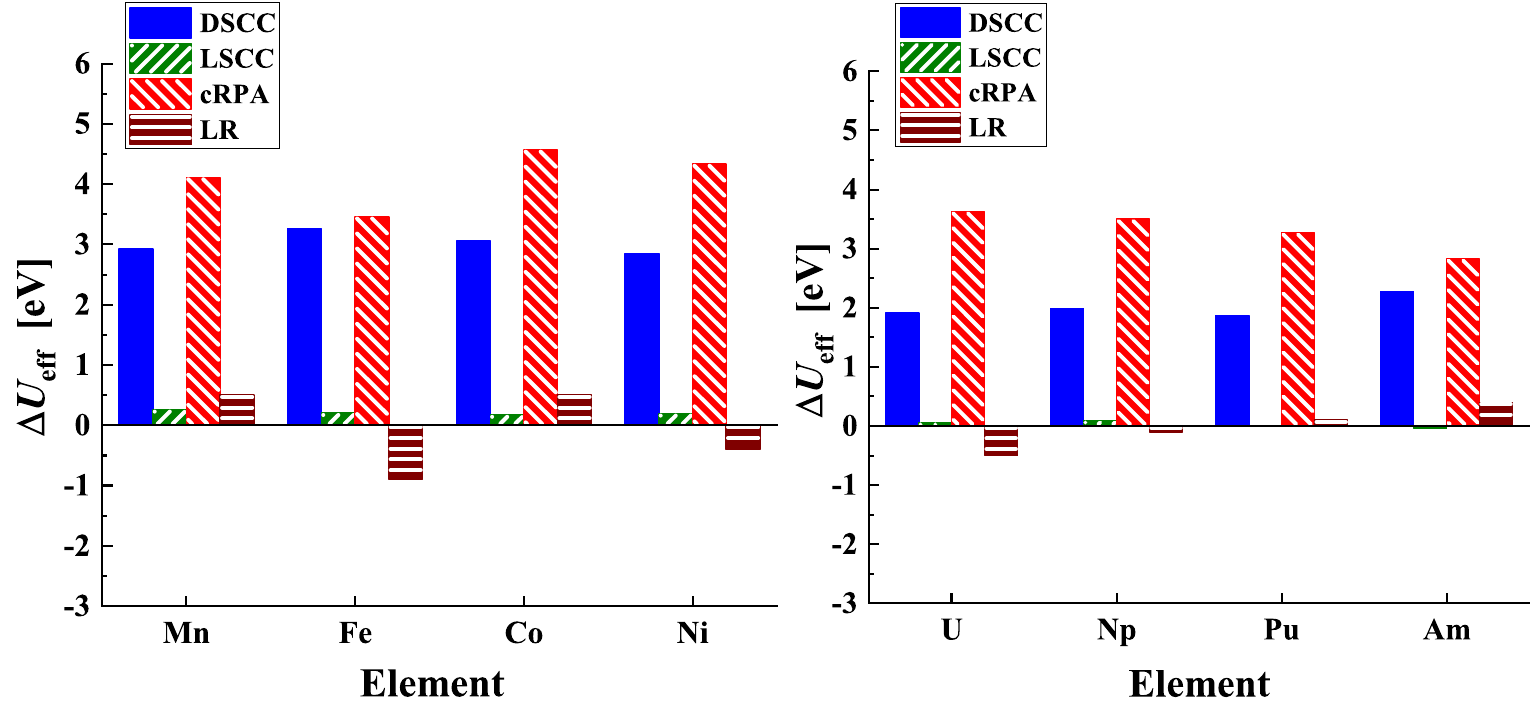}
\caption{The $U_{\rm eff}$ difference between metals and oxides ($\Delta U_{\rm eff}$) for the same element determined from DSCC, LSCC, cRPA and LR approaches. For transition-metals Mn, Fe, Co and Ni, $U^{\rm cRPA}_{\rm eff}$ derived by Shih et al. \cite{Shih2012}, $U^{\rm LR}_{\rm eff}$ derived by Nakamura et al. \cite{Nakamura2006}. For actinide metals U, Np, Pu and Am, $U^{\rm cRPA}_{\rm eff}$ derived by Amadon et al. \cite{Amadon2016,Amadon2018}, $U^{\rm LR}_{\rm eff}$ derived by Qiu et al\cite{Qiu2020}.}\label{fig:delta}
\end{figure*}

Research has suggested that $d/f$ electrons are more localized in oxides than in metals\cite{Shih2012,Moore2009,Lander1994}, i.e. $U_{\rm eff}$ are not as strong in the metals as in their oxides. We present the $U_{\rm eff}$ difference between metals and oxides ($\Delta U_{\rm eff}$) in Fig.\ref{fig:delta}. Both the DSCC and cRPA approaches reflect this feature well, while the differences of LSCC and LR are much smaller. Due to the smaller $U_{\rm eff}$ values provided by cRPA for metals\cite{Shih2012,Amadon2016,Amadon2018,Liu2023}, the differences are larger than for DSCC. Both DSCC and cRPA are dielectric screening methods, but in cRPA the dielectric function is constructed by the more computationally demanding random phase approximation. The DSCC approach shows similar performance to cRPA for different chemical environments. Therefore, it can also serve as an efficient alternative method for simulating on-site Coulomb interactions.

\section{Conclusion}
\label{sec:Con}

In this work, we presented the doubly screened Coulomb correction approach for strongly correlated materials. The DSCC method adopts the dielectric model, which considers both static dielectric and Thomas-Fermi screening mechanisms, to obtain on-site electron interactions self-consistently. We applied the DSCC approach to simulate the electron-structure properties including band gaps and magnetic order for a range of $d/f$-electron systems and compared with other theoretic and experimental results. Both $E_{\rm gap}$ and $E_{\rm AFM-FM}$ from DSCC are in good agreement with experiments (MAE 0.42 eV and 8.5 meV, respectively), which is comparable to conventional hybrid functionals, but  at a much lower computational. Compared with the LSCC method that only considers the Thomas-Fermi screening, the DSCC method has a better performance in the band gap, which indicates the importance of considering static dielectric screening.

Additionally, DSCC can properly describe the differences in correlation strength between different chemical environments, which is similar to the performance of the widely used cRPA, with a lower cost. It also proposed DSCC as an effective tool to simulate on-site Coulomb interaction parameters for more sophisticated methods such as DFT+DMFT and DFT+Gutzwiller.

Overall, despite the pragmatic performance of DSCC, there is still room for improvement. For example, work is in progress on the implementation of corrections for both the metal $d/f$-orbitals and the O-$p$ orbitals in strongly correlated metal oxides. While we only consider the on-site electronic interactions here, our approach can be easily extended to a framework considering inter-shell interactions. The Coulomb interaction could be accurately described by further considering the frequency dependent model dielectric function.


\section{Acknowledgement}
We thank Hong Jiang,  Yuan-Ji Xu,  Hua-Jie Chen,  Xing-Yu Gao, Ming-Feng Tian for helpful discussions. This work was supported by the National Key Research and Development Program of China (Grant No. 2021YFB3501503), the National Natural Science Foundation of China (Grant Nos. U2230401, U1930401, 12004048, and 11971066) and the Foundation of LCP. We thank the Tianhe platforms at the National Supercomputer Center in Tianjin.

\section{Author Declaration}
\subsection*{Conflict of Interest}
The authors have no conflicts to disclose.
\subsection*{Author Contributions}
\textbf{Bei-Lei Liu}: Conceptualization (equal); Investigation (equal); Writing - original draft (equal); Writing - review \& editing (equal).
\textbf{Yue-Chao Wang}:
Supervision (equal); Conceptualization (equal); Investigation (equal); Writing - original draft (equal); Writing - review \& editing (equal).
\textbf{Yu Liu}:
Supervision (equal); Conceptualization (equal); Writing - original draft (equal); Writing - review \& editing (equal).
\textbf{Hai-Feng Liu}:
Validation (equal); Writing - review \& editing (equal).
\textbf{Hai-Feng Song}:
Supervision (equal); Conceptualization (equal); Validation (equal); Writing - review \& editing (equal).

\section{Data Availability}
The data that support the findings of this study are available from the corresponding authors upon reasonable request.

\clearpage

\end{document}